\documentclass[
prb,
twocolumn,
superscriptaddress,
amsmath,amssymb,
aps,
]{revtex4-2}
\usepackage{appendix} 
\usepackage{graphicx}
\usepackage{dcolumn}
\usepackage{bm}
\usepackage[mathlines]{lineno}
\usepackage{braket}
\usepackage[T1]{fontenc}
\usepackage{comment}
\usepackage{xr-hyper}
\usepackage[ colorlinks, linkcolor=blue, anchorcolor=blue, citecolor=blue ]{hyperref}
\begin{document}
\title{Moiré flat bands and antiferroelectric domains in lattice relaxed twisted bilayer hexagonal boron nitride under perpendicular electric fields}
\author{Fengping Li}
\affiliation{Department of Physics, University of Seoul, Seoul 02504, Korea}
\author{Dongkyu Lee}
\affiliation{Department of Physics, University of Seoul, Seoul 02504, Korea}
\affiliation{Department of Smart Cities, University of Seoul, Seoul 02504, Korea}
\author{Nicolas Leconte}
\affiliation{Department of Physics, University of Seoul, Seoul 02504, Korea}
\author{Srivani Javvaji}
\affiliation{Department of Physics, University of Seoul, Seoul 02504, Korea}
\author{Jeil Jung}
\email{jeiljung@uos.ac.kr}
\affiliation{Department of Physics, University of Seoul, Seoul 02504, Korea}
\affiliation{Department of Smart Cities, University of Seoul, Seoul 02504, Korea}

\date{\today}
\begin{abstract}
Local interlayer charge polarization of twisted bilayer hexagonal boron nitride (t2BN) is calculated and parametrized 
as a function of twist angle and perpendicular electric fields through tight-binding calculations on lattice relaxed geometries
Lattice relaxations tend to increase the bandwidth of the nearly flat bands, where widths smaller than $\sim 1$~meV are expected for $\theta \leq 1.08^{\circ}$ for parallel BN/BN alignment, and for $\theta<1.5^\circ$ for the antiparallel BN/NB alignment. Local interlayer charge polarization maxima of $\sim$2.6~pC/m corresponding to interlayer electron density differences of $\sim 1.3 \times 10^{12} \rm cm^{-2}$ are expected at the AB and BA stacking sites of BN/BN aligned t2BN in the long moire period limit for $\theta \ll 1^{\circ}$, and evolves non-monotonically with a maximum of $\sim$3.5~pC/m at $\theta = 1.6^{\circ}$ before reaching $\sim$2~pC/m for $\theta = 6^{\circ}$.
The electrostatic potential maxima due to the t2BN moiré patterns are overall enhanced by $\sim 20\%$ with respect to the rigid system assuming potential modulation depths of up to $\sim 300$~mV near its surface. 
In BN/BN aligned bilayers the relative areas of the AB or BA local stacking regions can be expanded or reduced through a vertical electric field depending on its sign.
\end{abstract}

\maketitle

\section{\label{sec:level1} Introduction}
Twisted bilayer and multilayer graphene typically have their magic angles between $1^{\circ}$--$2^{\circ}$~\cite{bistritzer2011moire,park2022robust, harvard2021, Leconte2022} where the low energy bandwidths become nearly flat. When the Fermi level is tuned to lie within those bands they often give rise to correlated phases~\cite{cao2018correlated, xie2020nature, kerelsky2019maximized, xie2019spectroscopic, sharpe2019emergent,serlin2020intrinsic, wong2020cascade, zondiner2020cascade}, superconducting phases~\cite{cao2018unconventional, lu2019superconductors, yankowitz2019tuning}, and phase transition cascades as a function of carrier density~\cite{wong2020cascade, zondiner2020cascade}. 
Twisted gapped Dirac models~\cite{javvaji2020topological} that approximate some types of twisted semiconductor transition-metal dichalcogenides~(TMDs)
~\cite{yang2018origin, naik2018ultraflatbands, pan2020band, shimazaki2020strongly, an2020interaction, regan2020mott, zhang2020moire, wang2020correlated, tang2020simulation, li2021imaging} 
and twisted bilayer hexagonal boron nitride (t2BN)~\cite{xian2019multiflat} are expected to give rise to narrower low energy bands prone to correlation effects in larger twist angles than the ones from the graphene family. 
In particular, t2BN has drawn attention thanks to its interfacial local layer ferroelectricity 
whose charge polarization can be switched by perpendicular electric fields~\cite{yao2021enhanced, vizner2021interfacial, yasuda2021stacking, woods2021charge, rode2017twisted, moore2021nanoscale, ni2019soliton} by modifying the stacking geometry. 
%
Moreover, the moiré potential resulting from interlayer charge polarization offers an effective means to engineer the properties of adjacent 2D material layers~\cite{kim2024electrostatic}.
However, most tight-binding (TB) electronic structures currently rely on highly approximate models. For example, their TB hopping terms do not often distinguish the atomic species and assume a simplified distance dependence~\cite{walet2021flat}, or utilize interlayer hopping terms essentially similar to those of graphite~\cite{andjelkovic2020double}.
Moreover, the geometry for the atomic structure is often assumed rigid~\cite{xian2019multiflat,wang2020correlated}.


%

In this work, we report the atomic and electronic structure of t2BN using models informed by local density approximation (LDA) density functional theory (DFT) to systematically improve the theoretical framework to describe twisted bilayer and multilayer hexagonal boron nitride (hBN) systems. Specifically, we use exact exchange random phase approximation (EXX-RPA)-fitted force fields in \textsc{LAMMPS}~\cite{plimpton1995fast,plimpton2012computational} and calculate the electronic structure 
distinguishing the intralayer TB Hamiltonian modeled with distant neighbor hopping terms, and the interlayer coupling modeled through two-center~(TC) approximations
~\cite{javvaji_ab_2024} that we call the hybrid TC (HTC) model. 
More specifically, the TC model introduced for twisted bilayer graphene (tBG)~\cite{de2012numerical, leconte2022relaxation} is reparametrized to account for the species-specific interlayer interactions between B and N atoms. This HTC model reproduces more reliably the DFT band gap, the bandwidth, and layer-resolved local charge distributions for different twist angles. 

We will show that while lattice relaxations in t2BN tend to widen the band widths of the low energy nearly flat bands, they reduce below $\sim 1$~meV for twist angles smaller than $1.08^\circ$ and $1.5^\circ$ for the parallel BN/BN and anti-parallel alignments BN/NB respectively. 
These two types of alignments have different lattice relaxation behaviors and responses to a perpendicular electric field that alters the relative areas of different local stacking configurations. The external electric field has been simulated by adding layer-dependent onsite energy terms in the HTC model, inferred from the forces acting on the atoms during the \textsc{LAMMPS} calculation, offering an effective means to adjust the electronic properties of the system due to the changes in the moiré stacking domain sizes. 
Moreover, significant interlayer charge polarization domains in parallel BN/BN t2BN have been parametrized through a third-order harmonic expansion,
which allows modeling of the electrostatic potential induced in adjacent 2D material layers.

The paper is organized as follows. Section~\ref{methodologySect} introduces the methodology for the TB electronic structure parametrization and the atomic structure energy minimization calculations. Section~\ref{resultsSect} presents the lattice reconstruction effects and electronic structure results including band structures in the absence and presence of an electric field as well as charge transfer and induced moir\'e potential estimates, and in section~\ref{conclusionSect} we summarize our main findings.

\section{Methodology: systems and methods}
\label{methodologySect}

\subsection{Systems}

\begin{figure}[tbhp]
\centering
\includegraphics[scale=1.5]{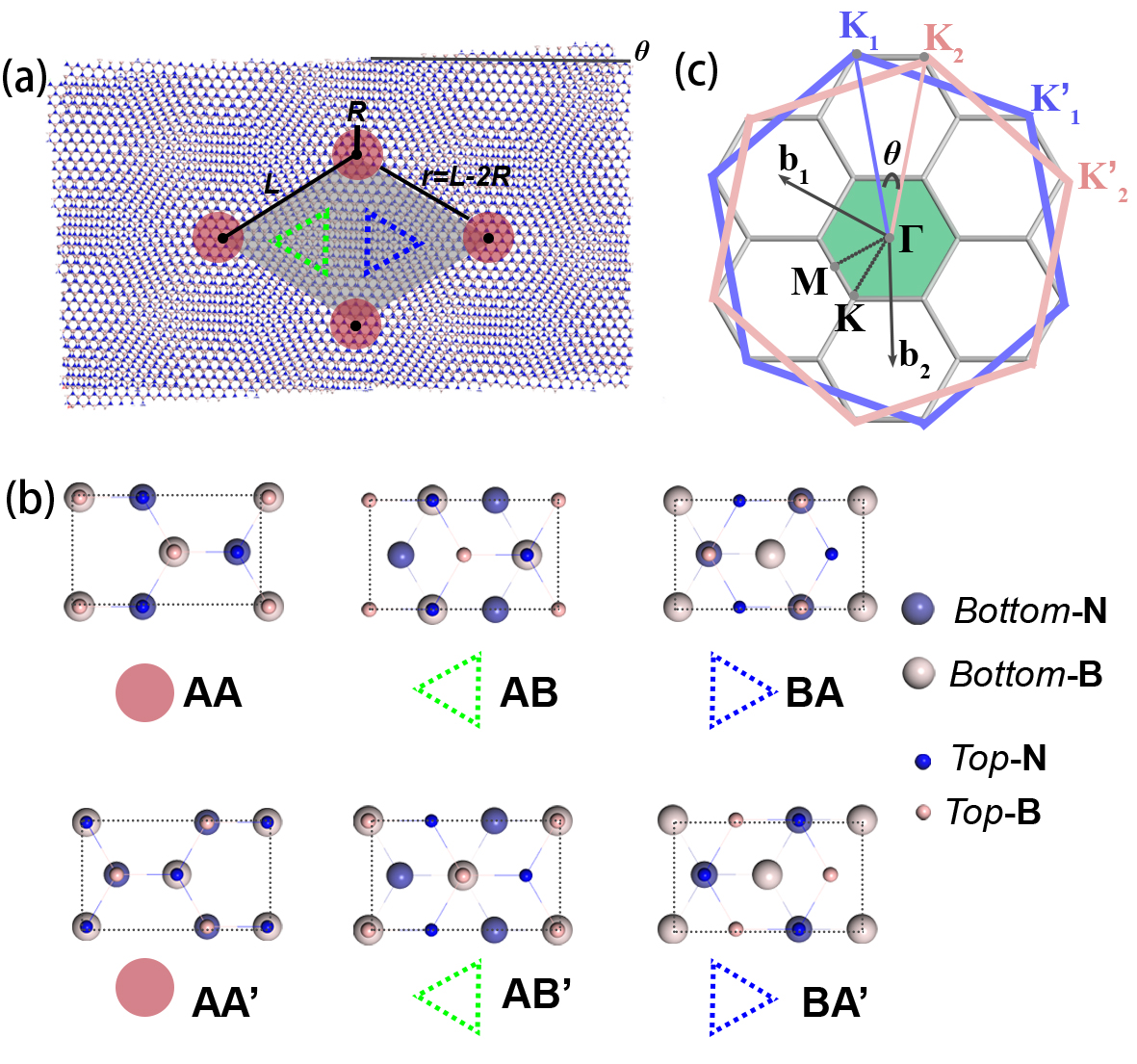}
\caption{Schematic representation of t2BN and local stacking geometries. 
(a) Real space moiré pattern for BN/BN with $\theta = 3.89 ^\circ$. The gray rhombus represents one unit cell of the moiré pattern. $L$ and $R$ correspond to the moiré length and the radius of the AA moiré stacking region. (b) Illustration of these high-symmetry stacking regions for BN/BN with AA-stacking rotation center (top row) and BN/NB with AA$^\prime$-stacking rotation center (bottom row). In both cases, the rotation center is located exactly where the sub-lattices of the bottom and top layers coincide.
Pink and blue balls denote the B and N atoms, respectively, and the top (bottom) layer is denoted by smaller (bigger) balls. (c) Schematic illustration of the moiré Brillouin zone (MBZ) in t2BN. The blue and pink hexagons denote the Brillouin zone corresponding to the top and bottom hBN layers, respectively. The smaller gray hexagons represent the MBZ formed by connecting the high symmetry points of the BZs of the constituent layers (e.g. $\mathbf{K_{1}}$ and $\mathbf{K_{2}}$). $\theta $ here matches the real space twist angle.
}
\label{fig:fig1}
\end{figure}

In the following, we describe the t2BN systems considered in our study. 
Fig.~\ref{fig:fig1}(a) illustrates the moiré pattern formed by twisting the two layers by a value of $\theta=3.89^\circ$, starting from parallel zero degree stacked BN/BN and choosing the rotation center at the first atomic site of AA stacking~\footnote{We choose the rotation axis to be at the first atom in the small unit cell that defines the local stacking.} 
in which boron and nitrogen atoms in one layer lie on top of the boron and notrogen atoms in the other layer. Similarly, the BN/NB moiré pattern can be constructed starting from anti-parallel BN/NB with the AA$^\prime$ local stacking axis as the rotation center, where the boron and nitrogen atoms in one layer lie on top of the nitrogen and boron atoms in the other layer (BN/NB). The high symmetry local stacking domain regions appearing in the moiré pattern of BN/BN and BN/NB systems are shown in Fig.~\ref{fig:fig1}(b). A gradual variation of stacking from AA ($\rm AA^{\prime}$) to AB ($\rm AB^{\prime}$) and BA($\rm BA^{\prime}$) domain regions can be recognized, reminiscent of the well-known moiré pattern strains in tBG moiré systems where the unstable high energy local stacking regions shrink and the low energy regions expand. In this work, we use the approach outlined in Ref.~\cite{leconte2022relaxation} to find commensurate cells for rotation angles from $21.78^\circ$ down to $1.08^\circ$. Fig.~\ref{fig:fig1}(c) displays the extended Brillouin zone for the moiré system resulting from a twist angle $\theta$ where ${\bm b}_{1}$ and ${\bm b}_{2}$ are the reciprocal lattice vectors. The electronic band structures in our calculations are shown for variations of the $\mathbf{\Gamma}$-$\mathbf{M}$-$\mathbf{K}$ path of the hexagonal MBZ region in reciprocal space while the local density of states is plotted at each of these high-symmetry points. 

\subsection{Methods}
The atomistic calculations of the relaxed lattice structures are combined with electronic structure calculations within the TB approximation. 
The atomic positions are relaxed using {\textsc{LAMMPS}}~\cite{plimpton1995fast,plimpton2012computational}, using the ExTeP (Extended Tersoff Potential) potentials for intralayer boron and nitrogen interactions~\cite{los2017extended}, while the interlayer interactions are described by the Dihedral Registry-Dependent Interlayer Potential (DRIP)~\cite{wen2018dihedral}, reparameterized using EXX-RPA data~\cite{leconte2017moire}, that can be used for both graphene and hBN interactions~\cite{leconte2022relaxation}. The first-principles calculations based on DFT performed with the software package \textsc{Quantum ESPRESSO} (QE) are interfaced with \textsc{Wannier90} to construct the Wannier functions~\cite{giannozzi2009quantum,mostofi2008wannier90,marzari2012maximally}. 
We use the norm-conserving Perdew-Zunger LDA pseudopotential~\cite{perdew1981self}, a $42\times42\times1$ k-points sampling density and $70$~Ry plane wave basis cutoff for the microscopic calculations of the bilayer with four atoms per unit cell. A $30$~$\text{\AA}$  perpendicular direction box size for each bilayer leaves enough vacuum space that suppresses the vertical interlayer coupling between the repeated cells~\cite{jung2014ab}.
The electronic structures and local densities for t2BN are obtained using a reparametrized hybrid TC model~\cite{de2012numerical,trambly2010localization,leconte2022relaxation}, hereafter indicated as the HTC model, where the intralayer hopping terms are characterized by considering up to six nearest neighbors for intra- and intersublattices~\cite{hbnsrivani,jung2013tight, jung2014accurate,carr2019derivation}, while we use the TC approximation for the interlayer hopping terms. 
 
For the HTC model parametrization, considering that the low energy conduction and valence bands (CB and VB) mostly stem from $p_z$ orbitals it is sufficient to use ten localized Wannier orbitals including four $p_z$ orbitals at the atomic sites and six $\sigma$ orbitals at the bonds. 
The calculated hopping amplitudes for intra-sublattice (B to B atom or N to N atom) and inter-sublattice (B to N atom) are listed in Table~\ref{tab:table1} 
and are closely similar to the ones in the model presented in Ref.~\cite{hbnsrivani}.
\begin{table}[tb!]
\caption{\label{tab:table1}
Six nearest neighbor intralayer hopping amplitudes in the top or bottom layer in the HTC model as defined in the text, where we take LDA values for the on-site energies of boron and nitrogen are $1.7666 $~eV and $-2.1843 $~eV, respectively. The indices in the first column refer to the $n$-th nearest neighbors as defined in Ref.~\cite{jung2013tight}.}
\begin{ruledtabular}
\begin{tabular}{rrrrc}
 &$t_\text{BB}$ ~&$t_\text{NN}$ ~& $t_\text{BN}$~ \\
\hline
$0$ & $1.7666$ & $-2.1843$ & $0.0000$ \\
$1$ & $0.0053$ & $0.1923$ & $-2.7205$ \\
$2$ & $0.0223$ & $0.0193$ & $-0.2102$ \\
$3$ & $-0.0483$ & $-0.0373$ & $0.0797$ \\
$4$ & $-0.0029$ & $-0.0027$ & $0.0047$ \\
$5$ & $-0.0033$ & $0.0000$ & $-0.0081$ \\
$6$ & $0.0002$ & $-0.0009$ & $0.0155$ \\
\end{tabular}
\end{ruledtabular}
\end{table}
The diagonal intralayer Hamiltonian elements, corresponding to row 0 in Table~ \ref{tab:table1}, are simulated taking into account the interlayer displacement-dependent Hamiltonian matrix element developed in Refs.~\cite{jung2013tight,jung2014accurate} whose sliding-dependent maps are summarized in Fig.~S.1 of the Supplemental Material
and whose expressions are reminded here as
\begin{equation}
H_{\alpha\beta}({\bm d}) = C_{0\alpha\beta}+2C_{1\alpha\beta}\operatorname{Re}\left[f({\bm d})\exp(i\varphi_{\alpha\beta})\right], \\
\label{eq:Onsite}
\end{equation}
with $\alpha\beta$ = B$_{\rm t}$B$_{\rm t}$, N$_{\rm t}$N$_{\rm t}$, B$_{\rm b}$B$_{\rm b}$ or N$_{\rm b}$N$_{\rm b}$ and
\begin{align*}
f({\bm d}) = \exp(-iG_1 d_y)+2\exp\left(i\frac{G_1 d_y}{2}\right)\cos\left(\frac{\sqrt{3}}{2}G_1 d_x\right)
\end{align*}
where ${d}_x$ and ${d}_y$ are the vector-components of the local stacking registry vector~\cite{jung2014ab, jung2015origin}
\begin{equation}
{\bm d}({\bm r}) = (\mathcal{ R}(\theta)-1){\bm r} + {\bm u}({\bm r})
\label{eq:eqD}
\end{equation}
where $G_1 = 4\pi/\sqrt{3} a$, taking $a$ as the reference lattice constant, where $\mathcal{ R}$ is the rotation operator acting on an atom at site ${\bm r}$ and where ${\bm u}({\bm r})$ captures the displacement of an atom due to lattice reconstruction.
The three parameters controlling this harmonic approximation (HA) are summarized in Table~\ref{tab:table2}.
\begin{table}[bt!]
\caption{\label{tab:table2}
The numerical values of $C_{0}$, $C_{1}$ and $\varphi$ in Eq.~(\ref{eq:Onsite}), where ${\rm B_{t}}$, ${\rm B_{b}}$, ${\rm N_{t}}$ and ${\rm N_{b}}$ denote respectively the top layer boron, bottom layer boron, top layer nitrogen, and bottom layer nitrogen atoms.}
\begin{ruledtabular}
\begin{tabular}{c|rrrrc}
 &${\rm B_{t}}$ ~&${\rm B_{b}}$~& ${\rm N_{t}}$ ~& ${\rm N_{b}}$~\\
\hline
$C_{0}$ & $ 1.7157 $ & $ 1.7157 $ & $-2.2560  $ &$ -2.2560 $\\
$C_{1}$ & $ 0.0099 $ & $ 0.0150 $ & $0.0150  $ &$ 0.0150 $ \\
$\varphi$ & $ 0.5437  $ & $ -0.5437 $ & $ -0.5437 $ &$-0.6525  $ \\
\end{tabular}
\end{ruledtabular}
\end{table}
For relaxed systems we account for variable intralayer bond length-dependence of the hopping terms through~\cite{pereira2009tight,leconte2022relaxation, hbnsrivani}
\begin{equation}
t_{\alpha\beta}({r}_{ij})=t_{\alpha\beta}({r}_{0,ij}) \exp\left[-2.45\left(\frac{{r}_{ij}-{r}_{0,ij}}{{r}_{0,ij}}\right)\right]
\label{bondlength}
\end{equation}
where $t_{\alpha\beta}({r}_{0,ij})$ are the hopping terms of the rigid lattice’s in-plane intralayer
interatomic distances ${r}_{0,ij}$ between $i$ and $j$ atoms and ${r}_{ij}$ is the relaxed distance. The $\alpha$ and $\beta$ refer to the sublattices in a single layer hBN. 
We used a lattice constant of $2.4795$~$\text{\AA}$\ for the reference rigid-structure DFT calculations to feed our model but the electronic structure corresponding to lattice relaxed structures which are close to the experimental lattice constant of $a = 2.504$~$\text{\AA}$~\cite{lynch1966effect} can be modeled through Eq.~(\ref{bondlength}).

After establishing the intralayer hopping terms, the interlayer hopping terms are determined using the Slater-Koster approximation~\cite{slater1954simplified}, which is expressed as
\begin{equation}
t_\text{inter} ({r}_{ij}) = {n^2_z} V_{p p \sigma}({r}_{ij})
 + \left(1-{n^2_z} \right) V_{p p \pi}({r}_{ij})
\label{TC}
\end{equation}
where ${n_z}=\ z_{ij}/{r}_{ij}$ with $ z_{ij}$ the z-coordinate of ${r_{ij}}$ and the coefficients used by the original two center approximation model~\cite{de2012numerical} for graphite are given by
\begin{equation}
V_{\mathrm{pp} \pi} ({r}_{ij})=\gamma_{0} \exp \left[q_{\pi}\left(1-\frac{{r}_{ij}}{a_{\rm BN}}\right)\right],
\label{vpppi}
\end{equation}

\begin{equation}
V_{\mathrm{pp} \sigma} ({r}_{ij})=\gamma_{1} \exp \left[q_{\sigma}\left(1-\frac{{r}_{ij}}{c_{\perp}}\right)\right].  
\label{vppsigma}
\end{equation}
We adopt this form for the hBN using 
$a_\text{BN} = a/\sqrt{3} = 1.43~$$\text{\AA}$ for the intralayer nearest neighbor distance of hBN, 
$c_{\perp}=3.261~$$\text{\AA}$ for the vertical interlayer distance,
and $\gamma_{0}=-2.7$~eV for the first nearest coupling term. 
The second nearest neighbor hopping term $\gamma_{0}^{\prime}$ is set to $0.1\gamma_0$, which is used to fix the ratio of $q_{\pi} / a$ such that 
\begin{equation}
\frac{q_{\sigma}}{c_{\perp}}=\frac{q_{\pi}}{a_{BN}}=\frac{\ln \left(\gamma_{0}^{\prime} / \gamma_{0}\right)}{a_\text{BN}-a}
\end{equation}
following the conventions and values from Ref.~\cite{de2012numerical} for tBG.
\begin{figure*}
\centering
\includegraphics[scale=0.65]{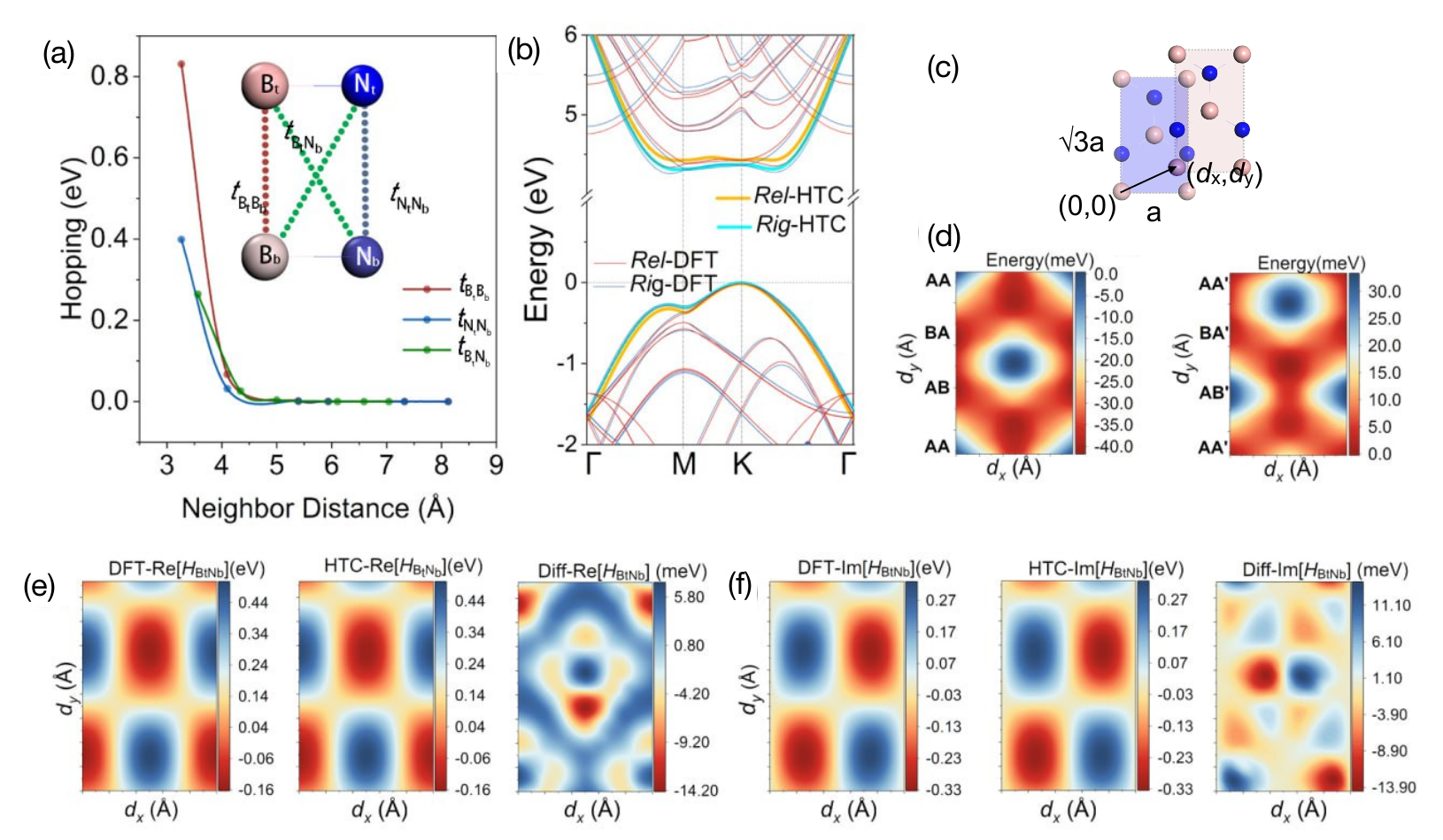}
\caption{\label{fig:fig2} 
Benchmark comparison of TB and DFT Hamiltonian and associated bands, as well as sliding-dependent energy landscapes, for the 4-atom commensurate cells. (a) In-plane distance-dependence of the nearest-neighbor interlayer hopping parameters used in the HTC model defined in the text. $t_{\rm B_{t}\rm B_{b}}$($t_{\rm N_{t}\rm N_{b}}$) denote the interlayer hopping terms between top B (N) atom and bottom B (N) atoms. The hopping terms for the interaction between top B or N atom with bottom N or B atoms are denoted as $t_{B_{t}N_{b}}$. (b) Comparison of the low energy bands from DFT and using the HTC model on BN/BN ($\theta=21.78^\circ$). The red (blue) real line and orange (cyan) lines represent the relaxed (rigid) configuration from DFT and HTC, respectively. The horizontal dashed line at $0$ eV is the Fermi level. (c) Illustration of the local registry vector ${\bm d} = (d_{x}$,$d_{y})$ showing the sliding between the bottom and top layer atoms in the same sublattice where the lattice is represented by the rectangle unit cell $a_\text{BN}$($1\times\sqrt{3}$) where $a$ is the lattice parameter. (d) The relative total energy landscape as a function of $d_x$ and $d_y$ sliding values for the same rectangle unit cell relative to AA and AA$^\prime$ stackings. (e) and (f) Real and imaginary parts of $\rm B_{t}\rm N_{b}$ interlayer matrix elements as a function of sliding based on AA stacking. DFT, HTC and Diff represent the tunneling data calculated by DFT Wannier representation, HTC model, and the difference between the DFT and the HTC model data, respectively.}
\end{figure*}
To improve on this generic model proposed originally for tBG, we reparametrize $\gamma_{1}$ in Eq.~(\ref{vppsigma}) to match the specific interactions that exist between B and N atoms.
This is done by matching the DFT interlayer tunneling at the $\mathbf{K}$ point~\cite{jung2014ab, leconte2022relaxation} defined as
\begin{equation}
\begin{aligned}
H_{s s^{\prime}}(\boldsymbol{K}: \boldsymbol{d})=\sum_{j_{s^{\prime}}} t_{i_{s} j_{s^{\prime}}}^{\mathrm{inter}} \exp \left[\mathrm{i} \boldsymbol{K} \cdot\left(\boldsymbol{d}+\boldsymbol{r}_{i_{s} j_{s^{\prime}}}\right)\right]
\end{aligned}
\label{tunnelingEq}
\end{equation}
where $s$ and $s^{\prime}$ refer to the two different bottom-layer sublattices $\alpha$ and $\beta$ and top-layer sublattices $\alpha^\prime$ and $\beta^\prime$. 
Together with Eq.~(\ref{TC}) this leads to the 
species-specific terms~\cite{hbnsrivani} given by 
$\gamma_{1}=t_{\text{BB}^\prime}=0.831$~eV, $\gamma_{1}=t_{\text{NN}^\prime}=0.6602$~eV, or $\gamma_{1}=t_{\text{BN}^\prime}=t_{\text{NB}^\prime}=0.3989$~eV, see Fig.~\ref{fig:fig2}(a).
Fixing the bottom layer sublattice by choosing one arbitrary $i_s$ site, we sum over all possible $j_{s^\prime}$ sites of sublattice $s^\prime$ in the top layer until the sum is converged. As $\mathbf{K}$ is a 2D vector, only the in-plane
components of the 3D distance vector $\bm{r}_{i_sj_{s^\prime}}$ contribute to the phase factor.
We can safely ignore the contributions proportional to $\gamma_0$ in this species-specific fitting procedure because the contribution from Eq.~(\ref{vpppi}) is negligible in Eq.~(\ref{TC}) for distances of the order of the interlayer distance $c_{\perp}$~\cite{de2012numerical}. Using BN/BN 4-atom commensurate cells for zero twist angle, we illustrate the DFT-tunneling maps, the fitted TB-tunneling maps, and their difference, for the real and imaginary parts of $t_{\rm{B_{t}N_{b}}}$ as a function of ${\bm d}$ in Fig.~\ref{fig:fig2}(e) and ~\ref{fig:fig2}(f). 
These displacement vectors ${\bm d}=(d_{x},d_{y})$ are illustrated in Fig.~\ref{fig:fig2}(c). The fitting of real and imaginary parts for the other terms $t_{\rm{B_{t}B_{b}}}$, $t_{\rm{N_{t}N_{b}}}$ and $t_{\rm{B_{t}N_{b}}}$ is illustrated in Fig.~S.2 of the Supplemental Material.
We note that for all these terms, both for their real and imaginary contributions, the maximum difference between DFT and TB modeling is less than a few percent and we can safely assume that our TB model offers an accurate parametrization of the main contribution to the interlayer tunneling.

We further benchmark the accuracy of our models by comparing the low-energy bands obtained from DFT and our HTC model for large twist angles. As shown in Fig.~\ref{fig:fig2}(b) for a $21.78^\circ$ BN/BN system, we find excellent agreement between the DFT and HTC $\pi$-bands especially at the $\mathbf{M}$ and $\mathbf{K}$ points both for relaxed and rigid configurations. The small differences near the $\mathbf{\Gamma}$ point are due to the $\sigma$-bands, which are not included in our HTC model.
To confirm the accuracy of our HTC model for different stacking configurations, we compare the low-energy bands obtained from DFT and HTC for six high-symmetry stackings. The results, shown in Fig.S.3 of the Supplemental Material,
reveal only minor differences at the lowest CB and the highest VB at the $\mathbf{K}$ point. Based on these benchmark calculations using commensurate cells, we can safely use the HTC model for the rest of the manuscript to calculate the electronic band structures and the local density of states in real space for the moiré systems.

To investigate the impact of an external electric field in the moiré cells of t2BN we introduce an additional correction proportional to the force introduced by an electric field ($F_z=qE_z$) on a given B or N atom.  
Using the electron charge $e$, the Bader charge~\cite{henkelman2006fast} values of $q = 0.82275|e|$ are used for B and $-q$ for N atoms, respectively~\cite{leconte2022relaxation}. The $z$ atomic positions relaxation under an external electric field using DFT (QE) and classical energy minimization simulations (\textsc{LAMMPS}) show both an approximately linear relationship and are in reasonable agreement with each other (see Fig.~S.4(a) of the Supplemental Material). 
Fig.~S.4(b) in turn displays a schematic of the atom position moving with the perpendicular electric field $E_z$ for AA and AA$^{\prime}$ stacking. 
respectively.
The site-resolved electric potential can be modeled through
\begin{equation}
V_i = z_i \frac{E_z}{\epsilon_r}
\end{equation} 
where $z_i$ is the z-position of atom $i$ within each relaxed structure. The origin point of the z-axis is defined such that $\braket{z} = 0$. We take relative dielectric constant of BN/BN to be $\epsilon_r = 2.7$ because it gives the best fit of the band gap as a function of perpendicular electric field in bilayer hBN. This value is comparable with the static perpendicular dielectric constant value of $\epsilon_r = 2.6$ reported in Ref.~~\cite{laturia2017dielectric} and somewhat smaller than $\epsilon_r = 3.4$ in Ref.~\cite{laturia2018dielectric} for bilayer hBN.  

\section{Results}
\label{resultsSect}
In the following, we analyze the atomic and electronic structure of t2BN in the absence and presence of a perpendicular electric field, together with the local ferroelectric charge distribution.

\subsection{Atomic structure}
The atomic structure of twisted BN/BN systems undergoes relaxations depending on the energetics of local stacking total energies at zero twist angle, in a manner similar to tBG. 
In panel Fig.~\ref{fig:fig2}(d), we show the energy landscape for zero twist angle BN/BN and BN/NB systems. The results reveal that for BN/BN systems, the AB and BA stackings have the lowest energy, whereas for BN/NB systems, the $\rm AA^{\prime}$ stacking is the most stable one due to the attractive interaction between the top B(N) atoms and the bottom N(B) atoms~\cite{constantinescu2013stacking,marom2010stacking}.
\begin{figure*}
\centering
\includegraphics[scale=0.7]{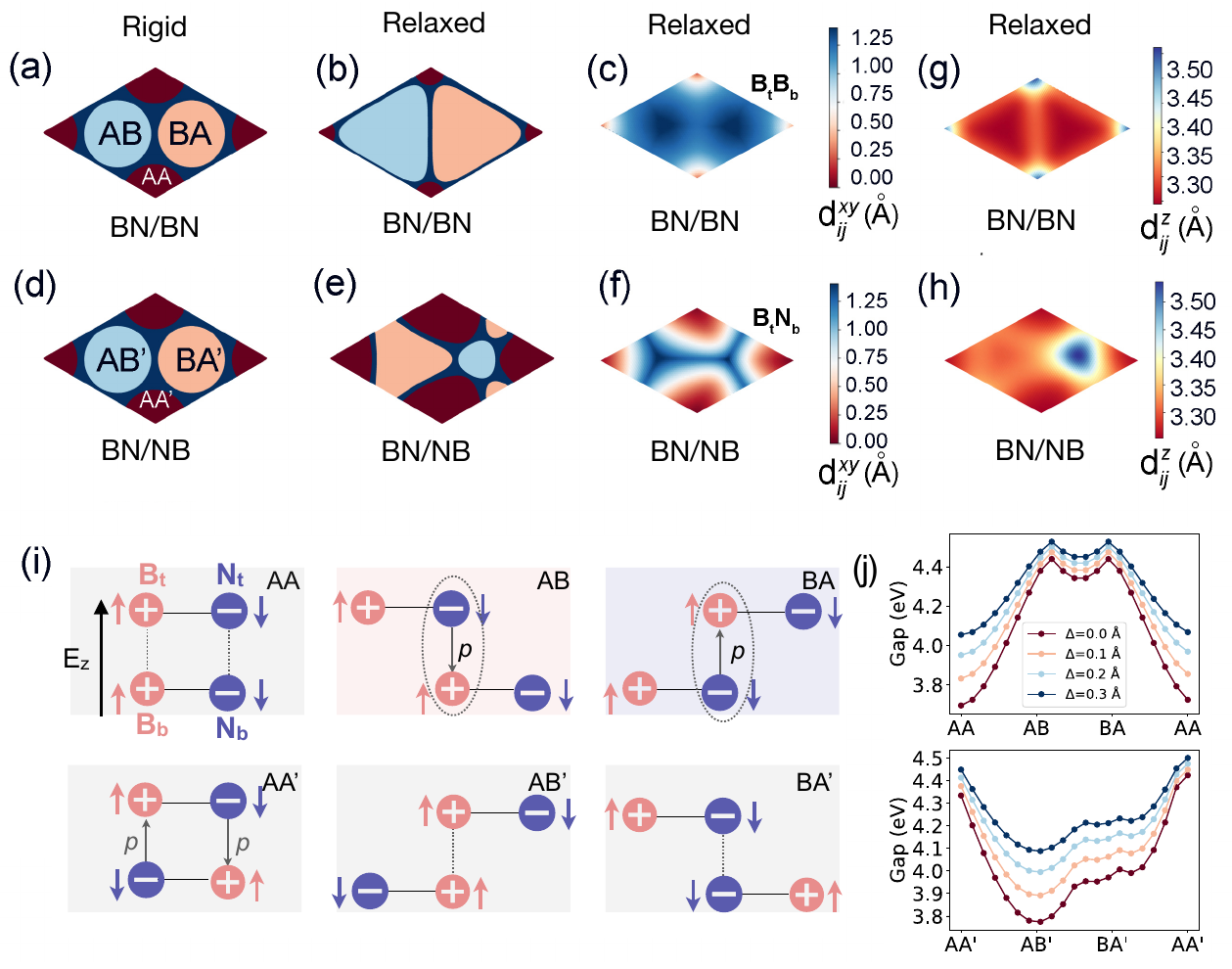}
\caption{\label{fig:fig3} Various representations of lattice reconstruction effects, schematics of charge transfer, and sliding-dependent gap estimates for the BN/BN and BN/NB systems, for a twist angle equal to $1.08^\circ$. (a) and (d) Stacking area maps of the different high-symmetry stacking regions determined by the in-plane displacement (also, see Fig.S.5 of the Supplemental Material)
for the rigid configurations of BN/BN and BN/NB, respectively. (b) and (e) Similar maps for the relaxed configurations of BN/BN and BN/NB. (c) and (f) The in-plane displacement distance $d^{xy}_{ij}$ for the relaxed BN/BN and BN/NB systems, where $i$ and $j$ indices refer to the closest pair of atoms in different layers belonging to the same sub-lattice, where $\rm B_{t}\rm B_{b}$ denote the top layer B atoms and bottom layer B atoms. (g) and (h) The out-of-plane disp lacement distance $d^{z}_{ij}$ for relaxed BN/BN and BN/NB configurations, respectively. (i) Schematic representation of the side views of the six high symmetry stackings under an external electric field. $\rm \bm{E}_z$ denotes the external electric field, $p$ denotes the dipole polarization between B and N atoms. The vertical motion directions of B and N atoms upon the application of $\rm \bm{E}_z$ are indicated by red and blue arrows. (j) The band gap as a function of stacking relative to AA (top) and AA$^{\prime}$ (bottom) stackings where the sliding path corresponds to the y direction in Fig. \ref{fig:fig1} (d). $\Delta$ denotes the variation in interlayer distance where $\Delta=0$ corresponds to the equilibrium interlayer distance.}
\end{figure*}
\begin{figure*}
\centering
\includegraphics[scale=0.65]{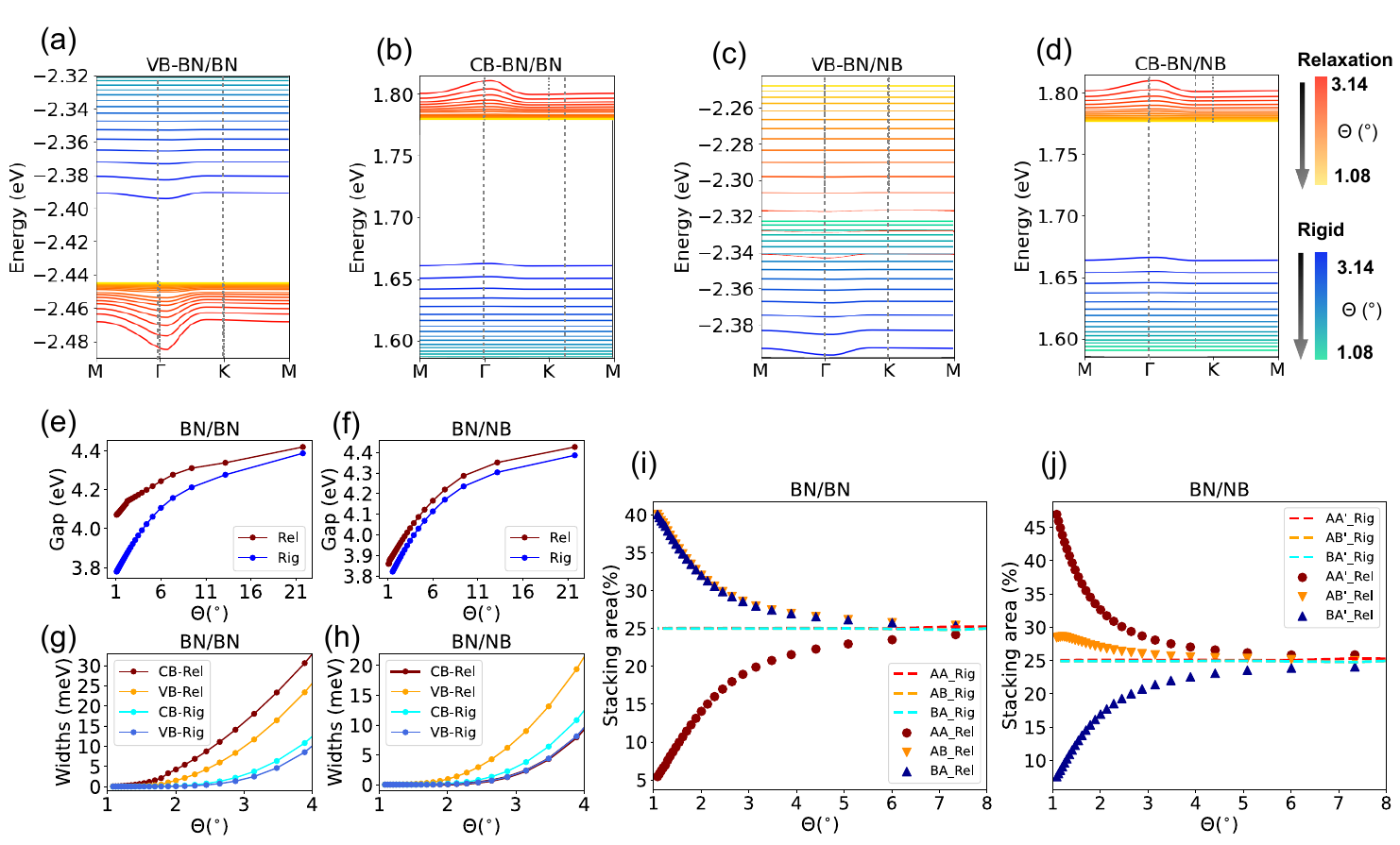}
\caption{\label{fig:fig4} Low-energy bands, band gaps, lattice-reconstructed stacking areas depending twist angle for the BN/BN and BN/NB systems. (a) and (b) Low energy CB and VB for various twist angles ranging from $3.14^\circ$ to $1.08^\circ$ in BN/BN. (c) and (d) Similar plots as (a) and (b) but for BN/NB. The red-to-yellow scale to denote the angle is used for the relaxed systems and the blue-to-green scale is used for the rigid systems. (e) and (f) Band gap as a function of twist angle for rigid (Rig) and relaxed (Rel) configurations in BN/BN and BN/NB, respectively. (g) and (h) Twist angle dependence of the VB and CB for rigid (Rig) and relaxed (Rel) moiré pattern in BN/BN and BN/NB. (i) The stacking area percentage with twist angles of relaxed and rigid configurations for BN/BN. (j) Similar plots for BN/NB.}
\end{figure*}
\begin{figure*}
\centering
\includegraphics[scale=0.28]{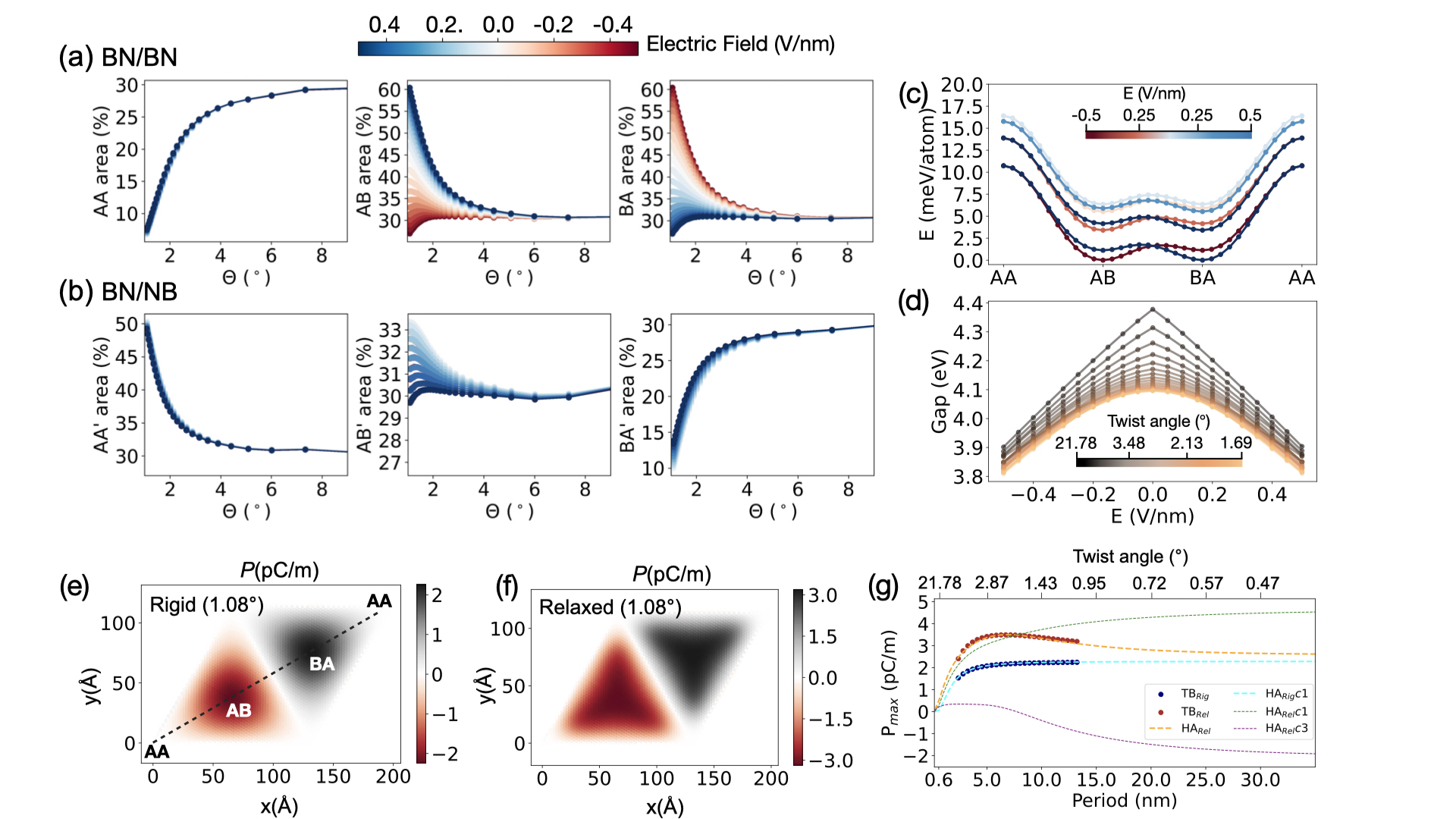}
\caption{\label{fig:fig5} Stacking regions and electronic properties under external electric fields for the BN/BN and BN/NB systems. (a) and (b) Percentage variation of stacking areas versus twist angle for BN/BN (AA, AB, and BA stacking regions) and BN/NB (AA$'$, AB$'$, and BA$'$ stacking regions) systems under a perpendicular electric field ranging from $-$0.5~V/nm to 0.5~V/nm. (c) Energy per atom dependence on sliding stacking relative AA stacking under external electric fields ranging from $-0.5$~V/nm to 0.5~V/nm. (d) Band gap of the BN/BN twisted system under various electric fields and different twist angles. (e) and (f) Electronic polarization distributions in rigid and relaxed $\theta=1.08^\circ$ BN/BN moir\'e systems. (g) The maximum and electronic polarization simulated by the TB model as a function of twist angle in the BN/BN system. Red and blue dots correspond to the relaxed and rigid configurations, respectively, while the corresponding HA is given with dashed lines. The individual first and third-order contributions to this HA for the relaxed system are presented as continuous dotted lines.}
\end{figure*}
We then assign a high symmetry local stacking to each atom in the moiré system following the approach illustrated in Fig.S.5 of the Supplemental Material.
This allows us to transform continuous in-plane displacement ${d}^{xy}_{ij}$ maps, as seen in  Fig.~\ref{fig:fig3}(c) and (f), to discrete maps in Fig.~\ref{fig:fig3}(a), (b), (d), and (e), which divide the relaxed domain regions into AA($\rm AA^{\prime}$), AB($\rm AB^{\prime}$)/BA($\rm BA^{\prime}$) stacking regions.
We use these discrete maps to have a simple and intuitive representation of the lattice reconstruction.

Continuous colormaps of the out-of-plane displacement ${d}^{z}_{ij}$, for both BN/BN and BN/NB systems, are shown in Figs.~\ref{fig:fig3}(g) and (h).
We notice a qualitative difference in the relaxation profile depending on the overall alignment sign. For BN/BN, the size of the AA-stacking region shrinks in favor of the other two local stackings, while the $\rm AA^\prime$ stacking in BN/NB has a larger domain region size after relaxation. The size of these different domain regions can be understood from the total energy landscape as a function of stacking shown in Fig.~\ref{fig:fig2}(d). 

Specifically, a more stable stacking gives rise to a larger domain size, while the unstable stacking shrinks to a smaller one.
In addition to the lattice reconstructions observed without an electric field, we also find that a vertical external electric field significantly impacts the system. Assuming that B and N atoms move in opposite directions due to their positive and negative effective charges, respectively, it makes sense to classify the interlayer interactions into four different contributions $\rm N_{t}\rm B_{b}$, $\rm B_{t}\rm N_{b}$, $\rm N_{b}\rm B_{t}$, and $\rm B_{b}\rm N_{t}$, where $t$ and $b$ indicate the top and bottom layers, respectively. This separation is illustrated in Fig.~S.4 of the Supplemental Material. 
%
By observing the evolution of the electric field-dependent interlayer distance $d^{z}_{ij}$ in Fig.~S.6 of the Supplemental Material
for both BN/BN and BN/NB systems, we can track the sizes of the six different local stacking regions of the BN/BN and BN/NB systems that respond differently depending on the sign of the electric fields, as illustrated in Fig.~\ref{fig:fig3}(i).
In this schematic, we illustrate the local interlayer charge polarization resulting from the overall positive charge accumulation on B atoms and negative charge accumulation on N atoms. The diagram demonstrates how the charge distribution within the interlayer region is influenced by the contrasting charges for B and N atoms. 

For BN/BN the layer charge polarization is achieved clearly for AB and BA stacking regions
where the interlayer dipole pointing along the $z$-direction is denoted by $p$ in Fig.~\ref{fig:fig3}(i). 
The opposite sign of the dipoles for different local stacking AB and BA domains implies asymmetric out-of-plane response under opposite electric fields. 
In contrast, for BN/NB no local polarization effect exists in the $\rm AA^{\prime}$, $\rm AB^{\prime}$, and $\rm BA^{\prime}$ stacking configurations, which explains why the sign of the electric field does not alter the local out-of-plane deformation nor the local interlayer separation. This is shown more clearly in Fig.~S.6(c) and (d) of the Supplemental Material
where we show the role of the electric field in modifying the stacking regions for BN/BN and BN/NB configurations, examining a range of electric fields spanning from $-0.5$~V/nm to $0.5$~V/nm. 
Due to the reversed polarity of AB and BA stackings in BN/BN, as shown in Fig~\ref{fig:fig3}(i), their respective sizes evolve in opposite ways for positive and negative electric fields. Conversely, for BN/NB, the different local stackings have almost no interlayer polarization, and result in nearly the same domain sizes and stacking registry maps regardless of the applied electric field. In the next sections, we relate these changes in the local stacking registry with the behavior of the low-energy bands, including the changes in the band gap size under an electric field.

\subsection{Electronic structure of twisted bilayer BN}

Nearly flat low energy bands without magic angles are expected near charge neutrality in sufficiently small twist angle t2BN thanks to the presence of a band gap that reduces the group velocity of the electrons near the band edges~\cite{ochoa2020flat, zhao2020formation, walet2021flat, xian2019multiflat, javvaji2020topological}.
Here, we revisit the electronic structure of this system with systematically improved atomic and electronic structure models to explore a wider range of twist angles, layer alignments, and perpendicular electric fields.
Specifically, as our HTC model distinguishes the hopping terms between different B or N atomic species, we study the 
BN/BN and BN/NB alignments in their relaxed and rigid configurations, in the presence and absence of an electric field.
We refer to Fig.~S.6(a) and (b) of the Supplemental Material for real-space local stacking distributions in different commensurate twist angles for BN/BN and BN/NB. 
In Fig.~\ref{fig:fig3}(j) we show the local band gap for zero twist systems for different interlayer sliding and distances. The vertical distribution of the positively and negatively charged boron and nitrogen sites can create a finite local interlayer dipole that leads to interlayer charge differences that we explore later. As expected, from an atomic structure point of view, the local stacking distributions undergo changes upon lattice relaxations that tend to expand the local stacking areas that are energetically more stable. 

In particular, we investigate the impact of twist angle and lattice reconstruction on the behavior of the low energy VB and CB. We present their evolution as a function of twist angle in Fig.~\ref{fig:fig4}(a)-\ref{fig:fig4}(d) where we compare the results of rigid and relaxed geometries. 
For rigid geometries, we use a fixed interlayer separation of $c_{\perp}$ = 3.261~$\text{\AA}$ as in Eq.~(\ref{vppsigma}).
We observe that the band gaps decrease for twisted systems after relaxation 
as shown in Figs.~\ref{fig:fig4}(e)-\ref{fig:fig4}(f).
As expected, the changes are more substantial for small twist angles when the moiré pattern strain profiles become more important. 
In the BN/BN case, we observe that the VBs shift to a lower energy region and the CBs shift to a higher energy region after relaxation. In contrast, in the BN/NB case, the CBs and VBs shift upwards in energy upon relaxation compared to the rigid case. 

To further understand the electronic properties in relation to the twist angle, we present in Figs.~\ref{fig:fig4}(g) and \ref{fig:fig4}(h) the variations of the bandwidths as a function of the twist angle for both rigid and relaxed superlattices in our two moiré patterns. The VB and CB that form the band gaps reach their maximum flatness below $1.08^\circ$ and $1.5^\circ$ degree for BN/BN and BN/NB, respectively. The reduced bandwidths of the VB and CB can be correlated with an increase in the size of the domain regions that contribute to those bands. 
Indeed, to check the wave function distribution on VBs and CBs, 
the electron density distribution at high symmetry points of $\mathbf{M}$, $\mathbf{\Gamma}$ and $\mathbf{K}$ are plotted in Fig.~S.7 of the Supplemental Material.
We show that the states at those high symmetry points distribute both at B and N atoms, near the AA stacking sites for BN/BN both for VB and CB states, while for BN/NB, the valence band states accumulate mainly at the N atoms near BA$^{\prime}$ stacking and CB states mainly accumulate at the B atoms near AB$^{\prime}$ stacking~\cite{zhao2020formation}. Thus, this distribution of electron density, in conjunction with the stacking area percentage allows us to understand the bandwidth variation.
In Fig.~\ref{fig:fig4}(i), the angle-dependence of stacking areas is presented for the BN/BN configuration. This diagram illustrates the decrease in the AA stacking region, which can be associated with the increased valence and conduction bandwidths shown in Fig.~\ref{fig:fig4}(g). Conversely, in the BN/NB configuration shown in Fig.~\ref{fig:fig4}(j), the reduction in the BA$^{\prime}$ region is associated with the increased valence bandwidth while the expansion of the AB$^{\prime}$ region to the decreased conduction bandwidth.

In the following, we consider the effect of applying a perpendicular electric field as a means to further control the structural and associated electronic properties.

\subsection{Electric field}
The electric field dependence of the atomic structure for the BN/BN and BN/NB systems is shown in Fig.~\ref{fig:fig5}(a) and (b) through the stacking area evolution in a BN/BN system. The local stacking regions gradually shrink or expand with varying twist angles. Using the conventions of stacking area definition outlined in Fig.~S.5 of the Supplemental Material,
the AA stacking reduces down to about 10\% of the total area, while AB/BA stackings almost make up 45\% in BN/BN system without external perpendicular electric field. 
When an external perpendicular electric field is introduced, the AA stacking region remains mostly unaffected. On the contrary, the AB and BA regions show an inverse trend for their relative ratios with respect to the total area. Specifically, the BA stacking area expands to 60\%, while the AB stacking area contracts to 25\% when subjected to a $0.5$~V/nm electric field, for small twist angles of around $\sim 1^\circ$.
The expansion and shrinking of the AB and BA stacking area ratios depend on the direction of the applied electric field.
These changes in local domain sizes stem from the impact of the electric field on the site potentials within specific stacking regions, as illustrated in Fig.~S.4(b) of the Supplemental Material.

Notably, B and N atoms move in opposite directions in response to an electric field due to their opposite charge.
This behavior can be understood by examining the energy dependence of the different stacking geometries when a vertical external electric field is applied, as shown in Fig.~\ref{fig:fig5}(c).
The AB and BA stacking configurations demonstrate opposite stability behaviors under the influence of positive and negative electric fields. 
%
%
Contrary to the BN/BN alignment we find a symmetric response to the electric field direction for the three representative local stacking configurations. Likewise, the energy-dependent sliding behavior of stacking can elucidate this relative insensitivity to an electric field. Indeed, the AA$^\prime$, AB$^\prime$, and BA$^\prime$ configurations demonstrate consistent stability under the influence of both positive and negative electric fields, see Fig.~S.4(c) of the Supplemental Material.
%

Our results show that the electric field can modify the overall interface interlayer ferroelectricity in t2BN by changing the size of the different stacking domains that have opposite layer charge polarization. 
This is further illustrated in Fig.~S.8 of the Supplemental Material.
The panels (d) in Fig.~\ref{fig:fig5} show the corresponding electric field evolution of the band-gap size relaxed BN/BN structures at different twist angles. The gap sizes decrease upon applying an electric field. 
%
%

We note that the dependence of electronic structure on perpendicular electric fields has been studied for a variety of 2D heterostructures combining graphene and hBN~\cite{chen2019signatures,long2023electronic,chittari2019gate,zhu2022electric}, 
and for hBN bilayers~\cite{yasuda2021stacking,wang2022interfacial,woods2021charge}
where perpendicular electric fields play a role in modifying the local stacking.
Experimentally, it was shown that in-plane sliding can switch the interlayer polarization in bilayer transition metal dichalcogenides and the specific stacking domains have associated layer-resolved charge polarization corresponding to those local stacking domains~\cite{wang2022interfacial}.    
Our current calculations on t2BN show qualitatively similar trends. 
The observed behavior stems from the repulsive potential difference between B-B or N-N, which linearly correlates with the change in the amplitude of the bandwidths. As a result, the electric field mediated 
change of local stacking in hBN bilayers~\cite{yasuda2021stacking,wang2022interfacial,woods2021charge}
can be an interesting pathway for engineering the electronic and optical properties in layer-charge polarized moiré systems. 
%
%
%
\begin{figure}[t]
\centering
\includegraphics[scale=0.90]{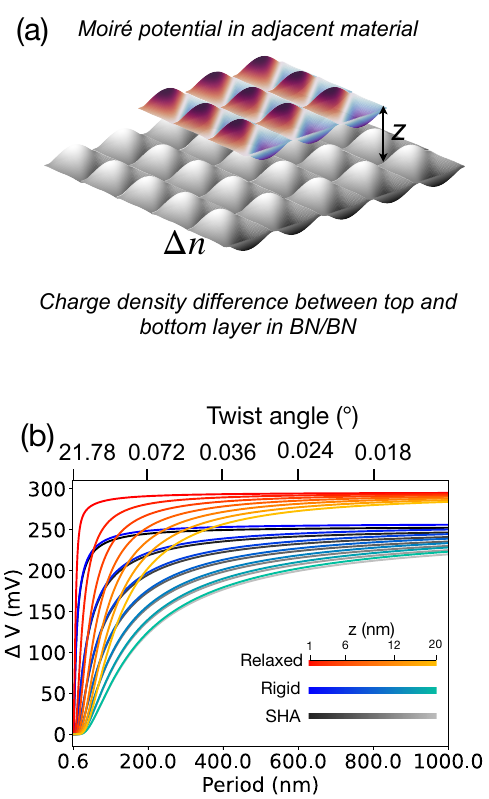}
\caption{Moiré potential observed in an adjacent material arising from its proximity with the polarized BN/BN moiré system.
(a) Schematic representation of the induced moiré potential in the adjacent material using a blue/red colormap. The gray map depicts the electron density difference between the top and bottom layers in the BN/BN moiré system.
(b) Moir\'e period dependence of the maximally induced potential for different vertical distances in relaxed and rigid BN/BN system with the gray gradient dashed lines representing the single HA (SHA) where S refers to using a constant electronic polarization of $2.25$ pC/m. We represent $\Delta V = V_{\rm max} - V_{\rm min}$ the 
the maximum potential modulation depths for different moire periods.}
\label{fig:fig6}
\end{figure}


\subsection{Interlayer polarization induced moir\'e potential}
\label{sec:chargeTransfer}

Following our discussion of the angle-dependence of the flat bands in our two moiré systems, we now shift our focus to the interfacial charge polarization, which has been observed in recent experiments~\cite{vizner2021interfacial,yasuda2021stacking,woods2021charge}. Interlayer electron density redistribution is expected in BN/BN moiré systems~\cite{walet2021flat,woods2021charge,yasuda2021stacking,vizner2021interfacial} but not in the BN/NB system. Indeed, for BN/NB alignment 
the interlayer charge polarization becomes zero at AA$^{\prime}$, AB$^{\prime}$ and BA$^{\prime}$ stacking region, which corresponds to zero intrinsic interlayer charge polarization as illustrated in Fig.~\ref{fig:fig3}(i). Hence, in the following electron density analysis, we focus only on the BN/BN system that has a significant interface domain polarization. 
Inverse electronic polarization can be observed between two stacking regions corresponding to the states located at either the AB or BA-stacking regions. 
In order to obtain the electron density difference and electronic polarization in the small twist angle regime, we fit the local interlayer electron density difference $n (\bm r)$ in BN/BN as a function of twist angle $\theta$
for each layer through a third-order HA for relaxed BN/BN, see Eqs.~(18)-(20) in Ref.~\cite{jung2015origin}, using here the real-space coordinate $\bm r=(x,y) $-dependence which reads,
after simplification, as
%
\begin{eqnarray}
\label{eq:harmonic}
 \Delta n({\bm r})  &=& 
 \sum_{j =1, 3}  2 c_{j}\left[ \sin(G_{j}  y) \right.   \nonumber \\
 &-&  \left. 4\cos(\sqrt{3} G_{j}  x /2) \sin (G_{j}  y/2) \right] 
\end{eqnarray}
where $c_{1}$ and $c_{3}$ are the coefficients of the first-order and third-order HA. 
Considering that only $c_{1}$ terms play a significant role in the rigid BN/BN system, the first-order HA is used for the rigid BN/BN simulation. In contrast, both the first-order and third-order HA are considered in the relaxed BN/BN simulation due to their concurrent sizeable contributions (see Supplemental Material in Fig.S.9).
The carrier densities are given in electrons per unit $\rm cm^2$.
We use $G_1=4\pi/\sqrt{3}a_{m}$, and $G_{3} = 2 G_{1}$, where $a_{m}$ is the real space periodicity of the moiré pattern.
%
The total interlayer electron density difference for top layer coordinate $\bm r $, 
can be obtained 
by adding the contributions from B and N atoms calculated separately using

\begin{equation}
\Delta n(\bm r) = \Delta n_{\rm B}({\bm r}) + \Delta n_{\rm N}({\bm r})
\label{eq:deltaN}
\end{equation}
where the species-dependent electron density differences are calculated from the electron density at each sublattice in the top and bottom layers (see Eq.~\ref{eq:harmonic}) as 

\begin{equation}
\Delta n_{\rm B}({\bm r})  =  n_{\rm B_{t}}({\bm r})  -  n_{\rm B_{b}}({\bm r})
\end{equation}

and
\begin{equation}
\Delta n_{\rm N}({\bm r})  =  n_{\rm N_{t}}({\bm r})  -  n_{\rm N_{b}}({\bm r}).
\end{equation}
%
%
%
This separation is more convenient for our calculations considering the strongly distinct electronegativities of the B and N sites.

The $\theta$-dependence of the parameters $c_{1}(\theta)$ for interlayer electron density of the rigid system and $c_{1}(\theta)$ and $c_{3}(\theta)$ for relaxed BN/BN systems are illustrated in Fig.~S.10 of the Supplemental Material
and can be fitted using the exponential function
\begin{equation}
c_j(\theta) = \alpha e^{\beta \theta}+\gamma e^{\delta \theta^{2}} \\
\label{eq:cphi_density}
\end{equation}
whose individual fitting parameters $\alpha$, $\beta$, $\gamma$ and $\delta$ are listed in Table~\ref{table:cphi_density}.
We provide parameterization sets both for rigid and relaxed geometries. Moreover, charge density at each sublattice in the top and bottom layers for rigid and relaxed BN/BN also provided in Fig.~S.11 and the fitting coefficients are listed in Table I of the Supplemental Material. 
%
\begin{table}[bt!]
\caption{\label{tab:table00}
The $\alpha$, $\beta$, $\gamma$, and $\delta$ fitting parameters in units of $cm^{-2}$ for  $c_{\Delta_{\rm B/\rm N}, 1}$, $c_{\Delta_{\rm B/\rm N}, 3}$ in Eq.~(\ref{eq:cphi_density}) to find $\Delta n_{\rm B}$ and $\Delta n_{\rm N}$ areal density differences between top and bottom B and N atoms using Eq.~(\ref{eq:harmonic}), for both rigid (Rig) and relaxed (Rel) BN/BN configurations. 
The $1$ and $3$ subscripts refer to the first and third-order coefficients.
It is worth noting that the first-order HA proportional to $c_1$ cannot accurately fit small twist angles interlayer electron distributions in the relaxed model and we must include the third order $c_3$ term.
}
\begin{ruledtabular}
\def\arraystretch{1.1}%
\begin{tabular}{lccccc} 
$\mathrm{ Rig}$  & $\alpha$ & $\beta$ & $\gamma$ & $\delta$  \\
\hline
${c}_{\rm \Delta B, 1}$ & $0.00$ & - & $5.81\times10^{11}$ & $-0.020$ &  \\
${c}_{\rm \Delta N, 1}$ & $0.00$ & - & $1.11\times10^{12}$ & $-0.006$ &    \\
\hline \hline
 $\rm Rel$ &$\alpha$ &$\beta$& $\gamma$ & $\delta$   \\ \hline
${c}_{\rm \Delta B, 1}$ & $ 1.24\times10^{12} $ & $ -0.13$ & $-2.46\times10^{11}$  & $-0.41$   \\
${c}_{\rm \Delta N, 1}$ & $ 1.93\times10^{12} $ & $-0.11$ & $6.46\times10^{11}$  & $-0.58$  \\
${c}_{\rm \Delta B, 3}$ & $1.73\times10^{11} $ & $-0.31$ & $3.64\times10^{12}$  & $0.49$  \\
${c}_{\rm \Delta N, 3}$ & $-3.57\times10^{11}$ & $ -0.03$ & $ 1.42\times10^{12}$  & $-0.64$  \\

\end{tabular}
\end{ruledtabular}
\label{table:cphi_density}
\end{table}



%
The electronic polarization
\begin{equation}P(\textbf{r})=-e\Delta n(\textbf{r}) c_{\perp}(\textbf{r})
\end{equation}
where $e = 1.602 \times 10^{-19}$~C,
can be obtained from the local interlayer electron density difference $\Delta n$ between top and bottom layers and the interlayer distance $c_{\perp}$.
Fig.~\ref{fig:fig5}(e) and (f) show TB-simulated electronic polarization occurring at the interface for rigid and relaxed $\theta=1.08^\circ$ BN/BN.
The local stacking registry ${\bm d}=(d_{x},d_{y})$ dependent electronic polarization and interlayer electron density difference $\Delta n({\bm d})$ for 4-atoms bilayer commensurate cells is shown in Fig.~S.12 for calculations based on TB, DFT, and the first harmonic (HA) fitting for zero twist angle. We find a nearly perfect agreement between TB and DFT, and the Harmonic approximation (HA) fit to first-order shows a difference in the density smaller than 1\% near density maxima points.
We find a good HA fitting of the electronic polarization for finite twisted BN/BN systems in Fig.~\ref{fig:fig5}(g) where we show the maximum electronic polarization for both relaxed and rigid systems as a function of twist angle, where the lattice relaxation reveals a more pronounced variation in the electronic polarization with twist, which is likely due to the larger AB and BA stacking regions in the relaxed model.
For relaxed and rigid BN/BN, the electronic polarization can be converged to $2.6$ pC/m and $2.23$ pC/m for large moiré patterns, respectively, where the rigid electronic polarization is in good agreement with theoretical and experimental values ranging from $2.00$ to $2.25$ pC/m in Refs.~\cite{yasuda2021stacking,zhao2021universal,li2017binary} for AB-stacked bilayer h-BN.
The relaxed BN/BN reaches a maximal interlayer polarization of $3.5$ pC/m at about $2.13^\circ$ ($\sim 7.5$ nm) due to the increasing importance of the third order terms at small angles that accounts for lattice reconstruction effects (purple dotted line in Fig.~\ref{fig:fig5}(g)). 


This interlayer electronic polarization in the BN/BN moiré configuration leads to induced moiré potentials that can influence the properties of adjacent materials~\cite{kim2024electrostatic}. 
We thus calculate such induced electronic static potentials using~\cite{zhao2021universal} 
\begin{equation}
V(\mathbf{r},z) \approx \sum_{j=1,3} - {\rm sgn}(z)\frac{e \Delta n_{j}(\textbf{r}) c_{\perp}(\textbf{r})}{2\varepsilon_{0}}e^{-G_{j}|z|}\\
\label{eq:moire potential}
\end{equation}
where $\varepsilon_{0}$ stands for the vacuum permittivity,
and $n_{j}(\mathbf{r})$ is the electron density difference at $\bm{r}$ for the $j^\text{th}$ order of the harmonic approximation. $z$ represents the vertical distance measured from the middle of the t2BN to the adjacent material, as illustrated in Fig.~\ref{fig:fig6}(a). Fig.~\ref{fig:fig6}(b) shows the moiré period dependence of the maximum potential difference value of $\Delta V = V_{\rm max} - V_{\rm min}$ for vertical interlayer distances ranging from $1$~nm to $20$~nm, in which $n_{j}$ corresponds to the electron density difference at the AB stacking in each BN/BN moiré pattern. The polarization is determined using a first-order HA ($j=1$) simulated electronic polarization based on the electron density distribution obtained through Eq.~(\ref{eq:harmonic}) (indicated by the blue to green gradient lines), yielding $240$~mV at a $1$~nm vertical distance. In contrast, relaxed BN/BN, based on the first and third-order HA ($j=1,3$)
(depicted by red gradient lines), exhibits an increased induced moir\'e potential, reaching $296$~mV. This value closely aligns with experimental findings~\cite{woods2021charge,yasuda2021stacking,vizner2021interfacial}. We further illustrate with the SHA curve in Fig.~\ref{fig:fig6}(b) (represented by gray gradient dashed lines) that using a constant polarization~\cite{zhao2021universal} of $2.25$~pC/m instead of the twist-angle-dependent values from Fig.~\ref{fig:fig5}(g) is a reasonable approximation when considering the rigid configuration.

We note that the induced potential does not sizeably change anymore for moir\'e patterns with periods larger than $600$~nm, \textit{i.e.} twist angles smaller than $\sim 0.02^\circ$. Furthermore, the simulated relaxed moiré potential demonstrates near vertical distance independence for large moiré lengths, with variations smaller than $10$~mV. This closely matches the distance dependence of moiré potential for large moiré pattern estimated in experiments.
Based on the observations in this Section, we speculate that the moiré potential arising from t2BN may significantly enhance the versatility of moiré engineering, thus potentially expanding the functionalities of electronic and photonic layered materials.

\section{Conclusions}
\label{conclusionSect}

In this work, we have quantitatively obtained the local interlayer charge polarization in t2BN that gives rise to moiré antiferroelectric domains when local stacking changes in real space. 
This calculation is carried out through a tight-binding model, informed by DFT calculations where we model separately the intralayer contributions with up to six nearest neighbor hopping terms and the interlayer tunneling is captured through a systematically improved distance-dependent two-center approximation that distinguishes the different atomic species. 
We have carried out our TB calculations on real-space lattice rigid and relaxed geometries and have made quantitative estimates on the effects that lattice relaxations have on the electronic structure of the system, generally enhancing the gaps and widening the bandwidths. 
We have further studied the atomic and electronic structures in the presence of an external electric field that can be used to expand or shrink certain local stacking regions.
Finally, the interface electron density profiles of the moiré patterns have been quantified through parametrization employing third-order harmonic expansions, thus enhancing their adaptability for future research endeavors. Based on these electron density profiles, we explore the moiré potential induced in the adjacent material contacted by t2BN in the BN/BN configurations, indicating potential avenues for manipulating the properties of neighboring functional layers by leveraging the surface potential of a t2BN substrate.

\begin{acknowledgments}
This work was supported by the National Research Foundation of Korea (NRF) through grant numbers NRF2020R1A5A1016518 (F.L.) and RS-2023-00249414 (N.L.). We acknowledge computational support from KISTI Grant No. KSC-2022-CRE-0514 and by the resources of Urban Big data and AI Institute (UBAI) at UOS. D.L. and J.J. also acknowledge support by the Korean Ministry of Land, Infrastructure and Transport (MOLIT) from the Innovative Talent Education Program for Smart Cities.
\end{acknowledgments}

\bibliography{aps}
\nocite{*}
\end{document}